\def\year{2023}\relax
\documentclass[letterpaper]{article} 
\usepackage{aaai21}  
\usepackage{times}  
\usepackage{helvet} 
\usepackage{courier}  
\usepackage[hyphens]{url}  
\usepackage{graphicx}  
\usepackage{natbib}  
\usepackage{caption} 

\usepackage{hyperref}
\usepackage{graphicx,balance,multirow,xspace,subcaption,longtable}
\usepackage{makecell,amsmath,cleveref,algorithm,color,colortbl}
\usepackage[font={small}, textfont=md]{caption}
\usepackage[T1]{fontenc}
\usepackage[draft,inline,nomargin,index]{fixme}
\usepackage{booktabs}
\usepackage{enumitem}
\usepackage[noend]{algpseudocode}

\definecolor{Gray}{gray}{.85}

\PassOptionsToPackage{hyphens}{url}
\newcolumntype{P}[1]{>{\arraybackslash}p{#1}}
\newcolumntype{X}[1]{>{\centering\arraybackslash}p{#1}}
\usepackage{pifont}

\FXRegisterAuthor{dd}{add}{Daniyal}
\FXRegisterAuthor{rn}{arn}{Rishab}

\definecolor{linkColor}{RGB}{6,125,233}
\hypersetup{%
  pdftitle={How Auditing Methodologies Can Impact Our Understanding of
  YouTube's Recommendation Systems},
  pdfauthor={},
  pdfkeywords={},
  bookmarksnumbered,
  pdfstartview={FitH},
  colorlinks,
  citecolor=black,
  filecolor=black,
  linkcolor=black,
  urlcolor=linkColor,
  breaklinks=true,
  hypertexnames=false
}

\newcommand\clearrow{\global\let\rowmac\relax}

\newcommand{\parai}[1]{{\it \noindent #1 }}
\newcommand{\parait}[1]{{\it \noindent #1 }}
\newcommand{\para}[1]{{\bf \noindent #1 }}

\crefformat{section}{\S#2#1#3}
\crefformat{subsection}{\S#2#1#3}
\crefformat{subsubsection}{\S#2#1#3}

\renewcommand{\footnotesize}{\scriptsize}
\newcommand{\etc}{etc.}
\newcommand{\eg}{e.g.,\ }
\newcommand{\etal}{et al.\xspace}
\newcommand{\ie}{i.e.,\ }

\newcommand{\cf}{{\em Cf.}\ }

\newcommand{\Wfull}{$W_{\text{100pc}}$\xspace}
\newcommand{\Whalf}{$W_{\text{50pc}}$\xspace}
\newcommand{\Wquart}{$W_{\text{25pc}}$\xspace}
\newcommand{\Wten}{$W_{\text{10pc}}$\xspace}

\newcommand{\Iget}{$I_{\text{get}}$\xspace}
\newcommand{\Iclick}{$I_{\text{click}}$\xspace}

\newcommand{\Pleft}{$P_{\text{left}}$\xspace}
\newcommand{\Pright}{$P_{\text{right}}$\xspace}

\newcommand{\Dtop}{$D_{\text{top}}$\xspace}
\newcommand{\Dbottom}{$D_{\text{bottom}}$\xspace}

\newcommand{\Afull}{$A_{\text{full}}$\xspace}
\newcommand{\Acookies}{$A_{\text{cookies}}$\xspace}

\newcommand{\Aclear}{$A_{\text{clear}}$\xspace}

\newcommand{\Tniche}{$T_{\text{niche}}$\xspace}
\newcommand{\Tmain}{$T_{\text{main}}$\xspace}

\newcommand{\Sniche}{$s_{\text{niche}}$\xspace}
\newcommand{\Smain}{$s_{\text{main}}$\xspace}

\newcommand{\treeseta}[1]{$\mathcal{T}_{\text{#1}}$}
\newcommand{\treesetb}[1]{$\mathcal{T}^{\prime}_{\text{#1}}$}

\author {
    Sarmad Chandio,
    Daniyal Pirwani Dar,
    Rishab Nithyanand\\
}
\affiliations {
    \{sarmad-chandio, rishab-nithyanand\}@uiowa.edu\\
    mdar@cs.stonybrook.edu\\
}

\begin{document}

\title{How Audit Methodologies Can Impact Our Understanding of YouTube's Recommendation Systems}


%
%

\maketitle
\sloppy

\begin{abstract}

  Data generated by audits of social media websites have formed the basis of
  our understanding of the biases presented in algorithmic content
  recommendation systems. 
  As legislators around the world are beginning to consider regulating the
  algorithmic systems that drive online platforms, it is critical to ensure the
  correctness of these inferred biases. However, as we will show in this paper,
  doing so is a challenging task for a variety of reasons related to the
  complexity of configuration parameters associated with the audits that gather
  data from a specific platform.

  Focusing specifically on YouTube, we show that conducting audits to make
  inferences about YouTube's recommendation systems is more methodologically
  challenging than one might expect. 
  There are many methodological decisions that need to be considered in order
  to obtain scientifically valid results, and each of these decisions incur
  costs. For example, should an auditor use (expensive to obtain) logged-in
  YouTube accounts while gathering recommendations from the algorithm to obtain
  more accurate inferences?
  We explore the impact of this and many other decisions and make some
  startling discoveries about the methodological choices that impact
  YouTube's recommendations. 
  Taken all together, our research suggests auditing configuration compromises
  that YouTube auditors and researchers can use to reduce audit overhead, both
  economically and computationally, without sacrificing accuracy of their
  inferences. Similarly, we also identify several configuration parameters that
  have a significant impact on the accuracy of measured inferences and should
  be carefully considered.
\end{abstract}

\section{Introduction} 
\label{sec:introduction}

\para{Auditing content recommendation systems is becoming increasingly
important.}
As social media platforms and the algorithms they employ continue to have an
increasing impact on our socio-political realities, auditing them (accurately)
has become an increasingly important task for many reasons.
After all, these audits, often focused on algorithmic recommendation systems,
play a significant role in drafting effective regulation around online
platforms and algorithms \cite{whitehouse2022} and developing a better
understanding of the role of algorithms in political polarization
\cite{barbera2020social}, spread of misinformation \cite{hussein2020measuring},
and other societal behaviors.
For example, focusing on the YouTube platform, prior work has uncovered several
concerning aspects of the algorithmic recommendation systems such as its
propensity to create filter-bubbles \cite{Tomlein-2021}, recommend
age-inappropriate content \cite{papadamou-2019}, misinformation
\cite{Tomlein-2021, hussein2020measuring}, and even extremist content
\cite{altright, incels}.
However, these works often appear to contradict each other --- \eg prior work has
shown that YouTube recommendation systems cause a mainstreaming effect (\ie
promoting popular content over niche content) \cite{ledwich-2019} while also
showing its tendency to promote niche and extremist content \cite{altright}.
Formulating effective regulation and developing a meaningful understanding of
the impact of algorithms on society is challenging in these scenarios where
contradictory findings from algorithm audit studies are commonplace. {\em This
work (1) shows that auditing methodologies are one source for such contradictory
results and (2) suggests approaches to reduce their occurrence.}

\para{Conceptually, designing a recommendation algorithm audit is simple.}
Due to the opacity of the algorithms being audited, researchers rely on what is
referred to as the ``sock-puppet'' audit approach \cite{sandvig-2014}. Here,
the audit can be generalized into a simple three-step process. 

\parait{Create sock-puppets.} Sock-puppets or personas
    that aim to impersonate real human users are created. The goal is to use
    automation tools, typically web crawlers, to provide the underlying
    recommendation system a set of interactions from which it may learn certain
    characteristics about the sock-puppet. In the context of YouTube, this may
    involve having the sock-puppet load a set of videos (referred to as the
    training set) that provide a base from which the recommendation algorithms
    learn user behaviors and preferences.

\parait{Measure the recommendation tree.} In this step, a seed
    interaction that generates recommendations is performed by the sock-puppet.
    This set of recommendations forms the first layer of the recommendation
    tree. Often, the recommendations themselves are interacted with recursively to
    form a deeper tree of recommendations. 
    Applied to YouTube, this step involves providing the sock-puppet with
    a seed video from which all recommendations are gathered. This is followed
    by then loading the videos associated with each of these recommendations
    themselves to fill the recommendation tree. 

\parait{Hypothesis testing.} Finally, given a (set of) recommendation
    tree(s) associated with sock-puppets of different characteristics,
    hypotheses about the underlying recommendation algorithm are tested and
    inferences about them are made.

\para{In practice, algorithm audits are challenging and can force
methodological compromises.}
Although simple at first glance, there are several key decisions in each step of
the previously described process that are often overlooked. For example, when
conducting crawls to construct sock-puppets, researchers are faced with the
decisions of what videos to use as part of their sock-puppet training set, how
many videos to include in this training set, and what video to use as their
seed, amongst others. The uncertainty about the impact that each video might
have on the gathered recommendations makes these decisions challenging.
Complicating matters, even when rigorous and sound rationale are applied to the
above questions, are the high dollar and computational costs associated with
methodological rigor. 

\parait{Methodological compromises due to high dollar costs.}
Online platforms, including YouTube, make it difficult to automate the creation
of the number of accounts required for a meaningful audit --- they serve
CAPTCHAs to (or perform outright blocking of) web automation tools seeking to
create accounts and often require verified phone numbers for each account.
The costs of circumventing these challenges can be prohibitively high and
force compromises that may impact the validity of their inferences. For
example, researchers may simply associate each sock-puppet with a unique
browser (cookie) and bypass the difficulties (and high costs) associated with
obtaining verifiable phone numbers for each sock-puppet. However, such
circumvention is often done in the hopes that the accuracy of any inferences
drawn from the audit are not harmed --- \ie they operate on the assumption that
YouTube's recommendations treat logged-in users in the same way as
non-logged-in users with YouTube's cookie in their browser. 

\parait{Methodological compromises due to high computational costs.}
Further, crawling videos is computationally expensive and time consuming when
crawlers encounter large numbers of hour-long (or longer) videos. This,
combined with the need to gather large amounts of data for statistically sound
hypothesis testing, can require 1000's of hours of machine time for a single
audit. {This poses another dilemma: should one pay the high computational
costs associated with watching the entirety of each video and should all paths
of the recommendation tree be traversed to make valid conclusions?} Although the
alternatives of simply sampling sections of the tree and not watching videos to
completion are more tractable, it remains unclear if they have an impact on the
subsequent recommendations gathered by the crawl.

Simply put, \emph{there are currently no best-practices or guidelines for
sock-puppet-style audits on platforms such as YouTube}. In this paper, we
specifically focus on YouTube and seek to fill this gap by answering the
following research questions.

\para{RQ1. What is the relationship between sock-puppet training set,
    recommendation seed, and recommendation trees? (\Cref{sec:priors})}
    We begin by studying the impact that the training set and seed have on the
    recommendation trees they generate. 
    Specifically, we conduct an experiment in which we train four sets of
    sock-puppets using all combinations of two distinctly different seeds and
    training sets. We then analyze the recommendation trees they generate to
    understand how recommendations change with alterations to the seed and
    training set. 
    %

\para{RQ2. What is the impact of reducing dollar costs during
    audits? (\Cref{sec:dollars})} 
    We investigate the consequences of one of the most commonly observed
    cost-saving measures adopted by YouTube auditors --- avoiding the use of
    real YouTube accounts for each sock-puppet and instead relying on browser
    cookies to leak a sock-puppet's identity to YouTube. 
    We conduct this analysis by comparing the recommendation trees generated by
    four sets of sock-puppets that reflect commonly observed practices in audit
    research. These sets of sock-puppets are identical in every way except for
    their method of maintaining YouTube account `state'.
    %
    
\para{RQ3. What is the impact of reducing computational costs during
    audits? (\Cref{sec:computation})} 
    Finally, we consider the consequences of compromises that are associated
    with reducing computational costs. We specifically focus on the impact of
    time spent on each sock-puppet training video and the depth/breadth of
    recommendation tree exploration. 
    We do so by training sets of sock-puppets that ``watch'' videos to varying
    levels of completion and measuring the differences in their gathered
    recommendation trees. We then study the characteristics of the nodes
    sampled from all recommendation trees gathered in our study to identify
    differences in their properties based on their location in the tree.
    %


\section{Methodology} \label{sec:methodology}

In total, we conduct eleven experiments (\cf
\Cref{tab:methodology:experiments}) in which we alter specific audit
parameters. We use the gathered recommendation trees from these audits to 
identify the impacts of varying parameters.
In this section, we describe our audit configuration selections and analysis
methodologies.

\subsection{Configuration parameters}\label{sec:methodology:configurations}

\para{Sock-puppet training sets} (\Tniche, \Tmain).
In all 11 experiments, we begin by training our sock-puppets with the videos
contained in either \Tniche or \Tmain.
A previous study \cite{papadamou-2022} shows that 22 videos are enough to
personalize the YouTube video recommendations, whereas we decided to
with 32 videos\footnote{The full list of
videos in each training set is available at \url{https://osf.io/3j5u8/?view_only=c2e21b99dbc3470e86bc9e904b39e6d3}}
for each of our training sets.

\parait{The niche training set (\Tniche).} The videos in \Tniche were
    manually curated to represent fringe (\eg conspiracy theories) and
    relatively unpopular (lower number of views) content. The videos in this
    set were chosen from fringe subreddits such as {\em r/climateskeptics} and
    {\em r/theworldisflat}, amongst others. Videos in this set were advocating
    for the position associated with the subreddit topic (\eg pro flat-earth).
    On average, videos in \Tniche had received 25K views. 


\parait{The mainstream training set (\Tmain).} Each video in \Tmain was
    curated to cover the same topic as their niche counterpart, except that
    they were sourced from a YouTube search of the topic (\eg `flat earth
    debunked'). The most popular videos from the search results (in terms of
    views) were added to \Tmain. The videos in \Tmain
    represent the 'mainstream' views and advocate for the highly accepted world
    view of the topics (\eg earth is not flat). On average, videos
    in \Tmain had 5.9M views.

This approach of training set construction offers two sharply differing inputs
to the recommendation algorithms so that any effect of training set on
recommendations is measurable. 


\para{Recommendation tree seeds} (\Sniche and \Smain). 
Seed videos are the starting point from which recommendation trees are gathered
(\ie the root of the recommendation tree). 
Similar to our training sets, we used one of two seeds (\Sniche and \Smain)
which were selected based on the intuition that they would have sharply
differing impacts on the recommendation tree. 
The \Sniche video used in our experiments was a fringe political video on the
topic of illegal immigration with 7.1K views and the \Smain video was a very
popular mainstream video focused on `Slapgate' \cite{slapgate} with over 3.8M
views. 
The topics of the seed videos were intentionally chosen (1) to not overlap with
the any of the videos from the \Tmain or \Tniche so that effects from the
training sets could be distinguished from those of the seed, and (2) to not
overlap with each other to maximize any measurable differences between their
recommendation trees. 
Observing an absence of differences in recommendation trees generated from
\Sniche and \Smain would indicate that the seed has a marginal influence on the
observed recommendations.

\para{Account status} (\Afull, \Acookies, and \Aclear).
To improve our understanding about whether the recommendation algorithm works
differently when audits operate under different YouTube account assumptions,
we gather recommendation trees using four different types of account
assumptions. 
(1) {\Afull} represents audits in which crawlers are logged into freshly
    created YouTube accounts before training and recommendation gathering
    begins. This is representative of the ideal case where each sock-puppet has
    its own fresh YouTube account.
(2) {\Acookies} represents audits where crawlers are not logged in but
    maintain YouTube's cookies in their browser throughout the crawl. This is
    representative of the most common crawls observed in audit literature.  
(3) {\Aclear} represents audits where crawlers conduct crawls while logged in
    and clear their watch history before the same account is used for another
    crawl. This approach is used to allow account reuse by different
    sock-puppets. 

\para{Watch times} (\Wfull, \Whalf, \Wquart, \Wten). 
Training sock-puppets can be computationally expensive owing to the long
lengths of videos typically contained within the training sets. 
To understand whether videos in the training set need to be watched to
completion, we gather recommendation trees from four crawlers all configured
identically except that they watch each of the training videos to different
levels of completion before moving on to the next video. 
\Wfull, \Whalf, \Wquart, and \Wten watch videos to 100\%, 50\%, 25\%, and 10\%
of completion, respectively. 
%

\para{Interactions} (\Iget, \Iclick).
Programming crawlers to perform actual clicks on hyperlinks is a challenging
task due to difficulties with reliability. A commonly used alternative is to
instead obtain links by parsing the DOM and having the browser load the link of
interest. 
Unfortunately, the absence of actual clicks is also a signature used by common
bot-detection tools and may result in server-side differential treatment
\cite{singh-usenix, khattak, suleman-2020, brave}. 
We conduct an experiment to understand whether clicking on recommended videos
impacts subsequent recommendations. 
\Iclick represents an audit in which each crawler actually performs a mouse
click on videos to load them during the recommendation tree crawl. \Iget
represents an audit in which each crawler simply obtains the video's URL from
the DOM and instructs the browser to load that URL. 
%
%

\para{Breadth of exploration} (\Pleft, \Pright). 
YouTube's recommendations are dynamically loaded and recommendation options
often continue to appear while a user scrolls down the page. This increases the
width of the recommendation tree at each level. 
In our pilot tests, we observed that the maximum number of recommendations was
at least 40 for each video (and much higher in many cases). 
We conduct analyses on the videos that appear at the top of the recommendation
list during a recommendation tree crawl (\ie the left-most path in the tree
denoted by \Pleft) and those that appear at the bottom of the recommendation
list (\ie the right-most path in the tree denoted by \Pright). 

\para{Depth of exploration} (\Dtop, \Dbottom).
Finally, we consider the importance of performing deep crawls on measured
characteristics of the recommendation tree. 
We do this by analyzing the characteristics of all videos observed after just
loading the seed video (\ie the $1^{st}$ level in the recommendation tree
denoted by \Dtop) and comparing them with the characteristics of all videos
observed at the $10^{th}$ level of the tree (\ie the bottom of \emph{our}
gathered recommendation trees denoted by \Dbottom).
%

\begin{table}
\centering
\small
\setlength\tabcolsep{3pt}
\begin{tabular}{cclcc}
\toprule
  \multicolumn{1}{c}{\bf Question}
& \multicolumn{1}{c}{\bf Parameter} 
& \multicolumn{1}{c}{\bf Configurations} 
& \multicolumn{1}{c}{\bf\#trees}
& \multicolumn{1}{c}{\bf\#videos} \\ 
\midrule
\multirow{2}{*}
{RQ1} & Training set & \Tmain, \Tniche    & 16 & 32K\\
      & Seed video   & \Smain, \Sniche    & 16 & 32K\\   
\midrule
\multirow{2}{*}
{RQ2} 	&   \multirow{2}{*}{Accounts}     & \Afull, \Acookies  & 8 & 14K\\
      	&          					   & \Afull, \Aclear & 8 & 13K\\   
\midrule
\multirow{5}{*}
  {RQ3} & \multirow{3}{*}{Watch Time} & \Wfull, \Whalf  & 8 & 15K\\
        &                             & \Whalf, \Wquart & 8 & 15K\\
        &                             & \Wquart, \Wten  & 8 & 15K\\
  \cline{2-5}
        & Interaction                 & \Iget, \Iclick  & 8 & 16K\\
        & Breadth          & \Pleft, \Pright & all* & 69K\\
        & Depth            & \Dtop, \Dbottom & all* & 35K\\
\midrule
\multicolumn{5}{c}{*Data from all trees were used in analysis for these parameters.}\\
\bottomrule
\end{tabular}
  \caption{{\bf Experiments.} The `Parameter' column indicates
  the parameter whose values were modified in each experiment and the
  `Configuration' column indicates the values assigned to the parameter in the
  experiment.
  The `\# trees' column indicates the total number of recommendation trees
  gathered for analysis and the `\# videos' column indicates the total number
  of recommendations observed in these trees.}
  \label{tab:methodology:experiments}

\end{table}

\subsection{Data gathering}\label{sec:methodology:gathering}


\para{Minimizing the influence of latent confounding variables.}
Recommendation trees are influenced by a large number of variables, some in
researchers' control (\eg our configuration parameters) and others not. 
In our study, we make a best-effort attempt to minimize these latent effects
with the following approaches.

\parai{Accounting for updates to the search index.} 
Due to large amounts of new content being created on YouTube, there are
continuous changes to the search index and recommendation candidate lists. 
Therefore, two crawls gathering recommendations at time periods that are far
spaced apart, may not be comparable due to vastly different recommendation
possibilities. 
We mitigate such impacts by synchronizing the crawls conducted in each
experiment such that for every crawler using one configuration to gather
a recommendation tree, there is another synchronized crawler using the
alternate configuration to gather the comparison recommendation tree. 
This synchronization is done at the node level --- \ie we ensure that each tree
arrives at the exact same node position in its respective recommendation tree
within, at most, a few seconds of its counterpart. 
%
Therefore trees gathered using alternate configurations of the same parameter
are comparable.

\parai{Accounting for distributed infrastructure and effects of geolocation.} 
As shown in prior work \cite{Hannak}, web servers may be distributed across
a wide region and servers in different locations or data centers may have
inconsistencies in their search indices or perform geo-specific recommendations. 
To mitigate these effects on our gathered trees, we conduct all our data
gathering experiments from the same location and use a static DNS entry for
YouTube which ensures that all our content requests and interactions with the
platform are served by web servers, at the very least, in the same region. 

\parai{Accounting for A/B testing.} Platforms have been known to conduct A-B
testing on their users while testing new features or algorithm updates
\cite{FB-ABtesting}. We make a best-effort attempt to mitigate
the effects of such testing by gathering data from \emph{at least eight
identical and synchronized crawls for each parameter tested in our study}.

\begin{figure}[t]
   \centering
   \includegraphics[width=.45\textwidth]{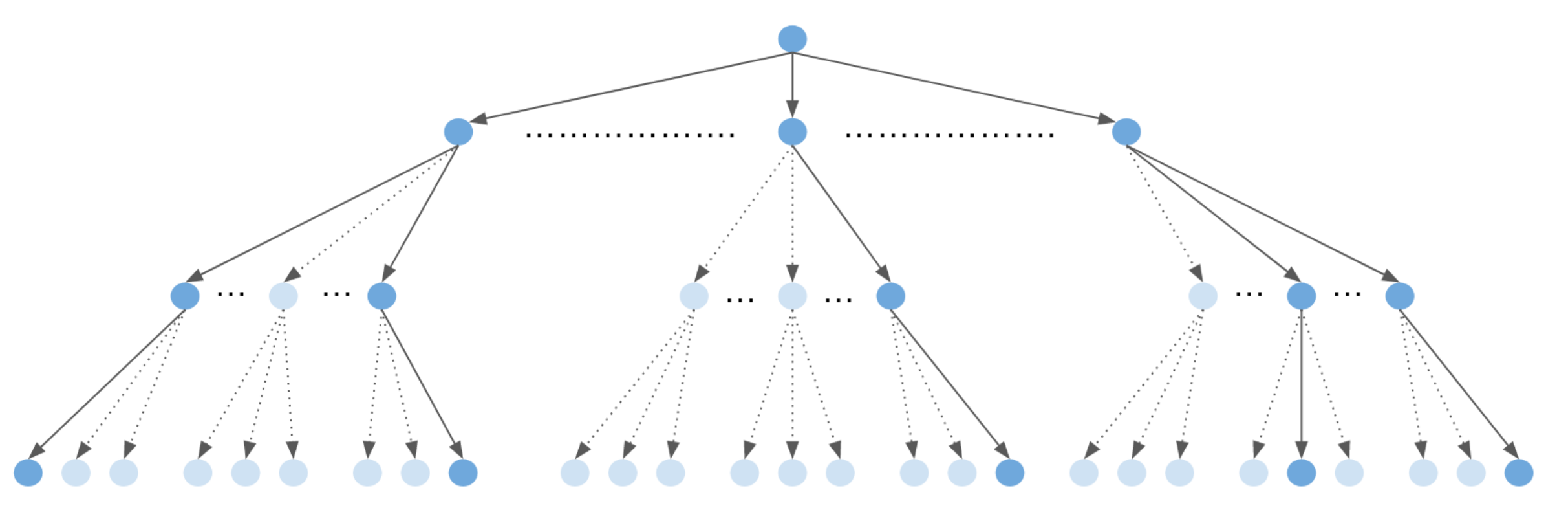}
   \caption{A recommendation tree generated based on 5 sock-puppets.
   Flat arrows shows the unique recommendation path taken by each sock-puppet.
   Starting from the seed, each node represents a video recommended
   by the parent node.
   }
   \label{fig:intro:recommendation-tree}
 \end{figure}

\para{Collecting recommendation trees.}
Once a sock-puppet has been trained and has a seed video, we begin exploration
of the recommendation tree. Unfortunately, complete exploration of
a recommendation tree is infeasible due to the need for one sock-puppet for
each configuration being tested for each tree being gathered for each path
being traversed. This is necessary due to the fact that prior watched videos
will impact future recommendations and therefore a sock-puppet can only perform
one-way (downward) traversals of the recommendation tree. 
Further, We are collecting at least 40 recommendations for each video.
Therefore, a recommendation tree of depth $n$ will have at least $40^n$ paths
from root to leaf node each needing a unique sock-puppet. 
%
In our traversals of the tree, we explored five unique paths --- the left-most
path (comprised of the first recommendation at each node), the right-most path
(comprised of the last recommendation at each node), and three pre-selected
paths from the middle (sampled with zipfian weights to account for a preference
for videos higher in the recommendation list).
We explore each of these paths simultaneously, using a unique but identically
trained, configured, and seeded sock-puppet dedicated to each, to a depth of 10
and record all recommendations along the way. We stitch these paths and
observations together to obtain a subset of the complete recommendation tree
upon which our analysis is conducted. We gather at least four 
such trees for each parameter configuration while ensuring
synchronization with alternately configured audits.
An example of such a tree is shown in \Cref{fig:intro:recommendation-tree}.
In this figure, each path represents the set of videos that form the sock-puppets,
nodes represent videos, and directed edges between any two nodes
($parent$, $child$) indicate that the $child$ was recommended after direct
interaction with the $parent$. The root node of this tree represents the
seed video used to generate the first set of recommendations.

\subsection{Recommendation tree characteristics} \label{sec:methodology:characteristics}
In our analysis, we focus on studying the popularity, channel diversity, and
topics of videos observed in a recommendation tree.
We select these parameters since platform audits often focus on them (or their
variations) to identify echo-chamber, rabbit-holing, or mainstreaming effects
caused by recommendation algorithms. 
%

\para{Popularity of recommended content.}
Popularity of recommended content, measured using video views as a proxy, can
capture the algorithm's tendency to recommend niche or mainstream content. We
record the distribution of views observed in recommended videos at each node.
A recommendation tree largely containing videos with low popularity at each
node suggests the tendency to recommend niche content for the associated
sock-puppet configuration.
Conversely, a tree largely containing videos with high popularity at each node
suggests the tendency to recommend mainstream content for the associated
sock-puppet configuration.
Significant differences in the within- and across-group differences between the
trees generated by two configurations would suggest that one of the two
configurations tends to more mainstream (popular) recommendations than the
other.

\para{Channel diversity of recommended content.}
Each video about a topic reflects the perspective of the channel that uploaded
it. Therefore, we use the diversity of channels in the trees as a proxy for the
range of perspectives provided by the recommended content. We measure this by
recording the entropy of channels observed in the recommended content at each
node.
A recommendation tree with a high entropy of recommended content at each node
indicates high recommendation diversity and suggests the absence of
a rabbit-holing effect.
Significant differences in the within- and across-group differences between the
trees generated by two configurations would suggest that one of the two
configurations tends to show less diverse recommendations than the other. 

\para{Semantic similarity of recommended content.}
We extract the titles and text descriptions associated with each video observed
in our tree. We then perform standard NLP preprocessing operations on these texts (\ie
tokenization, stop-word \& URL removal, removal of tokens observed in more than
50\% of our dataset, and lemmatization). 
We combine these processed texts for all videos observed at each node in a tree
and use it as a representation of the topics observed in the recommendations at
that node. 
In order to perform semantic comparisons between any two nodes, we use the
SpaCy document similarity method \cite{SpaCy} which uses a bag-of-words
approach to compute the cosine similarities between the average of all word
vectors in a text. While this approach is limited in that it does not capture
polarity of content (\eg anti-vaccine text and text from arguments refuting the
anti-vaccine texts will have high similarity), our manual validation found that
they captured the similarity of the topics in texts.
We also tested other semantic (Latent Semantic Indexing \cite{LSI}) and
lexicographic (Latent Dirichlet Allocation \cite{LDA}) approaches for measuring
similarity but found them to perform poorly at capturing topical similarity for
our dataset.
A pilot-study in which pairs of videos deemed to be very similar (by SpaCy's
docsim, LDA, LSA) were randomly sampled and their texts were manually evaluated
to verify similarity. Based on this, we identified Spacy's docsim as best.
Significant differences in the within- and across-group differences between 
trees generated by two configurations suggest that one of the two
configurations results in measurably different recommended topics.

\begin{table*}[ht]

\resizebox{\linewidth}{!}{%
\begin{tabular}{cc cllc cllc llc}
  \toprule
  \multicolumn{2}{c}{{\bf Parameters}} & 
  \multicolumn{4}{c}{{\bf Video Popularity (Views in millions)}} & 
  \multicolumn{4}{c}{{\bf Channel Diversity (Entropy in bits)}} & 
  \multicolumn{3}{c}{{\bf Content Semantics (Similarity score)}} \\ 
  
  Fixed & Varied &
  $\mu_{\text{views}}$   & Effect (95\% CI) & Effect (99\% CI) & $\mu_{\text{effect}}$ & 
  $\mu_{\text{entropy}}$ & Effect(95\% CI) & Effect (99\% CI) & $\mu_{\text{effect}}$ & 
  Effect(95\% CI) & Effect(99\% CI) & $\mu_{\text{effect}}$ \\ 
  \midrule

  \multirow{2}{*}{\Smain} & \Tmain & 
  7.15 & \multirow{2}{*}{{\bf[0.34, 1.33]}} & \multirow{2}{*}{{\bf[0.19, 1.49]}} & \multirow{2}{*}{{0.84}} & 
  3.63 & \multirow{2}{*}{[-0.16, 0.17]}   & \multirow{2}{*}{[-0.21, 0.22]}    & \multirow{2}{*}{0.00} & 
  \multirow{2}{*}{[-0.02, -0.00]} & \multirow{2}{*}{[-0.02, 0.00]} & \multirow{2}{*}{-0.01} \\

       & \Tniche & 
  4.94 &  &  &  & 3.49 &  &  &  &  &  &  \\  \midrule
  
  \multirow{2}{*}{\Sniche} & \Tmain & 
  4.32 & \multirow{2}{*}{{\bf [1.46, 2.19]}} & \multirow{2}{*}{{\bf [1.35, 2.30]}}  & \multirow{2}{*}{{1.82}} & 
  3.38 & \multirow{2}{*}{[-0.27, 0.19]}      & \multirow{2}{*}{[-0.34, 0.26]}       & \multirow{2}{*}{-0.04} & 
  \multirow{2}{*}{{\bf[-0.04, -0.02]}}       & \multirow{2}{*}{{\bf[-0.05, -0.01]}} & \multirow{2}{*}{{-0.03}} \\

  & \Tniche & 1.80 &  &  &  & 3.26 &  &  &  &  &  &  \\  \midrule

  \multirow{2}{*}{\Tmain} & \Smain & 
  10.71 & \multirow{2}{*}{{\bf [0.73, 2.31]}} & \multirow{2}{*}{{\bf [0.50, 2.56]}}   & \multirow{2}{*}{{1.51}} & 
  3.17  & \multirow{2}{*}{[-0.14, 0.14]}      & \multirow{2}{*}{[-0.18, 0.18]}        & \multirow{2}{*}{0.00} & 
  \multirow{2}{*}{{\bf [-0.04, -0.02]}}       & \multirow{2}{*}{{\bf [-0.04, -0.02]}} & \multirow{2}{*}{{-0.03}} \\

  & \Sniche & 7.78 &  &  &  & 2.97 &  &  &  &  &  &  \\ \midrule

  \multirow{2}{*}{\Tniche} & \Smain & 
  4.93 & \multirow{2}{*}{{\bf [2.68, 3.05]}} & \multirow{2}{*}{{\bf [2.62, 3.11]}}   & \multirow{2}{*}{{2.87}} & 
  4.02 & \multirow{2}{*}{{\bf [0.12, 0.45]}} & \multirow{2}{*}{{\bf [0.07, 0.51]}}   & \multirow{2}{*}{{0.28}} & 
  \multirow{2}{*}{{\bf [-0.05, -0.02]}}      & \multirow{2}{*}{{\bf [-0.05, -0.01]}} & \multirow{2}{*}{{-0.03}} \\

  & \Sniche & 1.72 &  &  &  & 3.44 &  &  &  &  &  &  \\ \bottomrule

\end{tabular}%
}
  \caption{Impact of changes caused by varying training sets (top 2 rows) and
  seeds (bottom 2 rows). Columns represent the mean node values observed in each
  group for a particular characteristic, the 95\% and 99\% confidence intervals
  for the measured effect sizes (\ie difference between within- and
  across-group differences; \cf \Cref{sec:methodology:comparison}), and the
  mean effect size. Values in bold indicate a statistically significant effect
  size at the corresponding confidence level.}
    \label{tab:rq1-results}
\end{table*}

\subsection{Comparison of Audit Configurations} 
\label{sec:methodology:comparison}
Each of our experiments result in two sets of recommendation trees --- one set
for each audit parameter configuration being tested (\eg \Afull and \Acookies).
Trees in each set are gathered in synchronization with each other. 
%
Given these sets of recommendation trees, we compute the {\em across-group}
(\eg between \Afull and its synchronized \Acookies tree) and {\em within-group}
(\eg between two synchronized \Afull (or, \Acookies) trees) differences along
the three dimensions described in \Cref{sec:methodology:characteristics}. We
describe this process below.

\para{Recording characteristics of a recommendation tree node.}
Let $n_{ij}$ denote a traversed node (\ie viewed video) located on path $P_i$
and at depth $j$ in a recommendation tree and $r_{ijk}$ denote the $k^{th}$
recommendation observed at $n_{ij}$.
At each node $n_{ij}$ we record: 
(1) a popularity scalar value $pop(n_{ij})$ = $\mu(views(r_{ij,1}) \dots
views(r_{ij,40}))$ representing the view counts of all observed
recommended videos at this node;
(2) a channel entropy scalar value $div(n_{ij})$ = $entropy(channel(r_{ij,1}),
\dots, channel(r_{ij,40}))$ representing the diversity of channels in the
recommended videos at this node; and 
(3) a document vector $doc(n_{ij})$ = $docvec(desc(r_{ij,1}), \dots,
desc(r_{ij,40}))$ which represents the document vector associated with the
video descriptions obtained from all recommended videos at this node.

\para{Comparing characteristics of recommendation trees.}
Given two recommendation trees $T$ and $T^{\prime}$, we compute the differences
in characteristics in a node position-dependant manner --- \ie we compute
differences in the popularity vector, channel entropy, and document vector for
each node position in $T$ and $T^{\prime}$.
These differences are computed as follows:
\[\delta_{pop}(T, T^{\prime}) = mean( [ \forall i, \forall j: pop(n_{ij}) - pop(n^{\prime}_{ij}) ]) \]
\[\delta_{div}(T, T^{\prime}) = mean( [ \forall i, \forall j: div(n_{ij}) - div(n^{\prime}_{ij}) ]) \]
\[\delta_{sem}(T, T^{\prime}) = mean( [ \forall i, \forall j: docsim(docvec(n_{ij}), docvec(n^{\prime}_{ij})) ]) \]
These values effectively capture the mean node-to-node differences between $T$
and $T^{\prime}$.
This node-to-node comparison is possible because all trees
gathered in our study traversed the same set of paths in the
recommendation tree.  
Maintaining this node position dependence in tree comparisons is important
because it handles differences in characteristics that might arise from the
position of a node in the recommendation tree. For example, comparing the top
recommendation at depth=1 from $T$ with the $40^{th}$ recommendation at
depth=10 from $T^{\prime}$ could result in misattributing differences in
tree characteristics that arise from changes in recommendation ranks to the
impact of an audit configuration change.

\para{Computing within- and across-group differences.}
Given two auditing configurations $\mathcal{C}$ and $\mathcal{C}^{\prime}$
which generate the sets of trees $\mathcal{T}$ and $\mathcal{T}^{\prime}$,
respectively, we compute: 
(1) the {\em within-group differences} as the distribution of differences in
characteristics observed between trees within $\mathcal{T}$ and
$\mathcal{T^{\prime}}$; and
(2) the {\em across-group differences} as the distribution of differences
observed between trees across $\mathcal{T}$ and $\mathcal{T}^{\prime}$. 
These are denoted by:
\[\Delta^{within}_{x}(\mathcal{T}) = [ \forall (T_i, T_j) \in (\mathcal{T} \times \mathcal{T}): \delta_{x}(T_i, T_j) ] \]
\[\Delta^{across}_{x}(\mathcal{T}, \mathcal{T}^{\prime}) = [ \forall (T_i, T_j) \in (\mathcal{T} \times \mathcal{T}^{\prime}): \delta_{x}(T_i, T_j) ] \] 
\[\forall x \in \{pop, div, sem\}\]
The within-group differences, computed over all trees generated
with identical audit configurations, allow us to establish a {\em baseline} of
characteristic variations caused by factors outside the control of the auditor
(\eg probabilistic recommendation algorithm, A/B testing, \etc). The
across-group differences showcase the differences caused by the change in audit
configuration {\em and} external factors.

\para{Quantifying the impact of audit parameter configurations.}
Given distributions $\Delta^{within}_{x}$ and $\Delta^{across}_{x}$ associated
with configurations ($\mathcal{C}$, $\mathcal{C}^{\prime}$), we use
bootstrapping with 1M samples \cite{diciccio1996bootstrap, efron1987better} to
create 95\% confidence intervals around the mean within- and across-group
differences. 
We also use these bootstrapped samples to compute 95\% confidence intervals
around the effect size --- \ie {\em the difference between the within- and
across-group differences bootstrap samples}.
Let [$CI_{lower}$, $CI_{upper}$] be the $N$\% confidence interval for the effect
size. We say that the effect is statistically significant at this confidence
level if and only if $(CI_{lower} \leq CI_{upper} < 0)$ or $(CI_{upper} \geq
CI_{lower} > 0)$ --- \ie {\em iff N\% of the bootstrapped samples have observed
effect sizes of the same polarity}. 
In our work, we report the 95\% and 99\% confidence intervals for effect sizes.
We also report the average effect size as the mean of all effect sizes observed
in the bootstrap samples.

%

\section{Training sets and seeds}\label{sec:priors}

\para{Experiment setup.} Our goal is to measure the impact of training sets and
seeds on the characteristics of recommendation trees generated by an audit.
To accomplish this, we gathered 32 recommendation trees from four different
audit configurations: eight trees each from an audit using \Tmain and \Smain,
\Tmain and \Sniche, \Tniche and \Smain, and \Tniche and \Sniche. 
We split each of these into two sets of four and refer to them as 
(\treeseta{main,main}, \treesetb{main,main}),
(\treeseta{main,niche},\treesetb{main,niche}),
(\treeseta{niche,main},\treesetb{niche,main}), and
(\treeseta{niche,niche},\treesetb{niche,niche}) respectively.
These trees were gathered in synchrony (\cf \Cref{sec:methodology:gathering})
in order to facilitate accurate within- and across-group comparisons (\cf \Cref{sec:methodology:comparison}). By
splitting each of our sets of eight trees into two sets of four, we avoid
reusing trees for testing multiple hypotheses.

\parait{Measuring impact of a training set change.}
To uncover the impact of the training set used in an audit on the
characteristics of recommendation trees, we compute the means, 95\% and 99\%
confidence intervals associated with the within-group differences, across-group
differences, and effect sizes (\cf \Cref{sec:methodology:comparison}) obtained
from two analyses:
(1) comparing \treeseta{main,main} with \treeseta{niche,main} --- \ie using the
same mainstream seed while varying the training set; and
(2) comparing \treeseta{main,niche} with \treeseta{niche,niche} --- \ie using the
same niche seed while varying the training set.

\parait{Measuring impact of a seed change.}
We repeat our methodology for the following analyses: 
(1) comparing \treesetb{main,main} with \treesetb{main,niche} --- \ie varying the
seed while using a mainstream training set focused on controversial topics; and 
(2) comparing \treesetb{niche,main} with \treesetb{niche,niche} --- \ie varying
the seed while maintaining a fringe and controversial training set.

\para{Results.} Our results are summarized in \Cref{tab:rq1-results}. 
In general, we find that altering the characteristics of the training set or
the seed {\em always} impacts the popularity of the videos observed in an
audit. This, however, is not the case for the channel diversity and semantics.
More specifically, our analysis yields the following insights.

\parait{There appears strong evidence of a `recency bias' in recommendations.}
Paying attention to the bottom two rows of \Cref{tab:rq1-results}, we see that
the effects of altering the seed from a niche video to a mainstream video are
nearly always statistically significant and of high magnitude, with only one
exception when channel diversity is recorded using \Tmain for training.
The (significant) effects on the popularity and entropy of recommended videos
are also higher than the effects observed on alterations of the training
set (top two rows).
The most notable effects of altering seeds are in the `popularity' dimension 
where the mean effect of switching a seed video from niche to mainstream
results in video recommendations that, on average, have 1.51M and 2.87M more
views when trained with \Tmain and \Tniche, respectively. 
We only find marginal (yet significant) changes in the semantics of recommended
videos, however --- \ie recommendations are between 1-5\% less semantically
similar after switching seeds from mainstream to niche.
This suggests that, independently of the training set used, the choice of seed
can drastically alter the characteristics of a recommendation tree and the
audit inferences.
Extrapolating this finding suggests that the {most recent video will have
an outsized impact on future recommendations.}

\parait{Channel diversity is not always dependent on the training set and
seed.} 
Our analysis shows that the channel diversity is largely unaffected by the
choice of training set and seed.
Only one exception occurs: when seeds are altered for a \Tniche training set
audit (\cf row four in \Cref{tab:rq1-results}). Here we see the effect of
switching from \Smain to \Sniche reduces the channel diversity by an average of
0.28 (entropy in bits) at each node. 
While it appears that this finding lends credence to the claims of the
algorithms rabbit-holing tendencies, it is important to note that this decrease
only appears when the audit has only interacted with fringe content (in the
training set and the seed). Given that the effect disappears when any other
interaction occurs, this finding could be explained by the small number of
creators addressing the topic of the niche content.
\begin{table*}[ht]

\resizebox{\linewidth}{!}{%
\begin{tabular}{c cccc cccc ccc}
  \toprule
  \multirow{2}{*}{{\bf Parameters}} & 
  \multicolumn{4}{c}{{\bf Video Popularity (Views in millions)}} & 
  \multicolumn{4}{c}{{\bf Channel Diversity (Entropy in bits)}} & 
  \multicolumn{3}{c}{{\bf Content Semantics (Similarity score)}} \\

    & $\mu_{\text{views}}$ & Effect (95\% CI) & Effect (99\% CI) & $\mu_{\text{effect}}$ & 
  $\mu_{\text{entropy}}$   & Effect (95\% CI) & Effect (99\% CI) & $\mu_{\text{effect}}$ &
  Effect (95\% CI) & Effect (99\% CI) & $\mu_{\text{effect}}$ \\\midrule
  
  \Afull & 9.20 & \multirow{2}{*}{[-0.90, 0.99]} & \multirow{2}{*}{[-1.20, 1.29]}    & \multirow{2}{*}{0.05}  &
   		   3.36 & \multirow{2}{*}{[-0.11, 0.31]} & \multirow{2}{*}{[-0.18, 0.36]}    & \multirow{2}{*}{-0.01}  & 
    			  \multirow{2}{*}{[-0.02, -0.00]} & \multirow{2}{*}{[-0.03, 0.00]} 	 & \multirow{2}{*}{-0.01} \\

  \Acookies & 7.72 &  &  &  & 3.57 &  &  &  &  &  &  \\
  \arrayrulecolor{black}\midrule

  \Aclear & 12.34 & \multirow{2}{*}{\bf{[1.82, 3.46]}} & \multirow{2}{*}{\bf{[1.54, 3.70]}} & \multirow{2}{*}{2.65} & 
  			3.55  & \multirow{2}{*}{[-0.03, 0.53]} 	 & \multirow{2}{*}{[-0.12, 0.62]}  	& \multirow{2}{*}{0.26} & 
  					\multirow{2}{*}{\bf{[-0.05, -0.01]}} & \multirow{2}{*}{\bf{[-0.05, -0.01]}} & \multirow{2}{*}{\bf{-0.03}} \\
	\Afull & 8.47 &  &  &  & 2.86 &  &  &  &  &  &  \\ \hline
\end{tabular}%
}

  \caption{Impact of changes caused by varying login status (row 1)
  and purging watch history (row 2). Columns represent the mean node values observed in each
  group for a particular characteristic, the 95\% and 99\% confidence intervals
  for the measured effect sizes (\ie difference between within- and
  across-group differences; \cf \Cref{sec:methodology:comparison}), and the
  mean effect size. Values in bold indicate a statistically significant effect
  size at the corresponding confidence level.}
    \label{tab:rq2-results}
\end{table*}

\para{Takeaways.}
Taken together, these results put a different perspective on YouTube's
recommendation system and the audits that study it. 
Not only do researchers need to pay particular attention to training and
seeding, but also must understand that their measurements of recommended videos
are heavily  dependent on the {\em most recent} nodes already traversed by
their sock-puppets. 
Specifically, it appears that the recency bias can lead to a single video
overwhelming the effects of a large number of prior videos --- thus impacting
the final inferences from the audit. 
Generally, we recommend that audit inferences (\eg presence of a mainstreaming
effect) are conditioned: (1) on the specific characteristics of the training
set and seed; and (2) on the specific strategies used to select nodes from
a recommendation tree.

\section{Dollar-Cost Saving Configurations} 
\label{sec:dollars}

\para{Experiment setup.}
In this section, we focus on understanding the impact of  commonly
used sock-puppet account management strategies on the recommendation trees
generated by them.

\parait{Measuring the effectiveness of cookie-based sock puppets.}
To find out the differences in cookie based sock puppets against real accounts,
we gathered four recommendation trees for \treeseta{full} and  
\treeseta{cookies} each. All the parametric configurations
for these two sets were kept identical except \treeseta{full} was using a logged in
profile while \treeseta{cookies} was not logged in, but was maintaining YouTube cookies.
Both \treeseta{full} and \treeseta{cookies} used the (\Tmain, \Smain) training set
and seed.

\parait{Measuring the effectiveness of clearing account history.}
To verify whether clearing account history does indeed purge the watch history effect
(i.e even after deleting watch history, user keeps getting similar recommendations),
we collected four recommendation trees for \treesetb{full} and 
\treeseta{clear} each. Both \treesetb{full} and \treeseta{clear} 
were using logged in profiles and were using (\Tmain, \Smain) training set
and seed. However, before collecting recommendations based on seed 
\Smain, watch history of \treeseta{clear} was deleted. \\
To gain insights into the measurable effects of different account
management strategies, we compute the means,
95\% and 99\% confidence intervals associated with within-group
differences, across-group differences, and effect sizes.
%

%

\para{Results.}
The results are summarized in \Cref{tab:rq2-results}. Our analysis yielded
two conclusive results. 

\parait{Audits do not need fresh accounts for each sock-puppet.}
First, focusing on the impact of changing between
a sock-puppet with a logged-in YouTube account (\treeseta{full}) and one
which only maintains its browser cookies (\treeseta{cookies}), we found that there were
no significant differences in any measured characteristics of their
recommendations. This presents significant
cost-saving opportunities that arise from being able to associate a
sock-puppet with a browser instance rather than having to navigate 
the barriers associated with automating account creation and
phone number verification.

\parait{The potential for account reuse by clearing history}
There is a significant difference in popularity and content semantics
for \treeseta{full} sock-puppets when compared with identically configured
and synchronized \treeseta{clear} sock-puppets, suggesting
that, by clearing history \treeseta{clear}
has purged the popularity-context and topic-context (picked up during training phase)
which \treeseta{full} still maintains. Simply put, by
clearing account history one might be able to reuse an account for
a large-scale study --- particularly where the popularity and content
semantics are being measured (\eg in audits quantifying 
mainstreaming and rabbit-holing effects). However, 
we do not make the claim that clearing watch history is equivalent
to getting a fresh account (a fresh account would mean Google doesn't have
any data stored for the profile at the back end, which we did not check for).

\para{Takeaways.}
These findings present an opportunity for auditors to save huge
dollar-costs involved in account creation and curation. We have shown
that a browser that maintains YouTube cookies is as good as YouTube
account. Furthermore, account re-use (after clearing
 history) is a viable option for auditors
studying the platform for its popularity and 
content semantics.
%
%
%
\begin{table*}[ht]

\resizebox{\linewidth}{!}{%
\begin{tabular}{cccccccccccc}
  \toprule
  \multirow{2}{*}{{\bf Parameters}} & 
  \multicolumn{4}{c}{{\bf Video Popularity (Views in millions)}} & 
  \multicolumn{4}{c}{{\bf Channel Diversity (Entropy in bits)}} & 
  \multicolumn{3}{c}{{\bf Content Semantics (Similarity score)}} \\ 

    & $\mu_{\text{views}}$ & Effect (95\% CI) & Effect (99\% CI) & $\mu_{\text{effect}}$ & 
  $\mu_{\text{entropy}}$   & Effect (95\% CI) & Effect (99\% CI) & $\mu_{\text{effect}}$ &
  Effect (95\% CI) & Effect (99\% CI) & $\mu_{\text{effect}}$ \\\midrule
  
  \Wfull & 7.69 & \multirow{2}{*}{[-0.86, 0.28]} & \multirow{2}{*}{[-1.04, 0.46]}        & \multirow{2}{*}{-0.29} & 
  3.74   & \multirow{2}{*}{[-0.23, 0.22]}        & \multirow{2}{*}{[-0.30, 0.29]}        & \multirow{2}{*}{0.00} & 
  \multirow{2}{*}{[-0.03, -0.00]}                & \multirow{2}{*}{[-0.03, 0.00]} 		 & \multirow{2}{*}{-0.01} \\

  \Whalf & 7.84 &  &  &  & 3.51 &  &  &  &  &  &  \\ 
  \arrayrulecolor{Gray}\hline

  \Whalf & 12.13 & \multirow{2}{*}{[-3.52, 1.53]} & \multirow{2}{*}{[-4.28, 2.34]} & \multirow{2}{*}{-1.03} & 
  3.21   & \multirow{2}{*}{[-0.41, 0.15]}         & \multirow{2}{*}{[-0.50, 0.25]} & \multirow{2}{*}{-0.13} & 
  \multirow{2}{*}{[-0.03, 0.00]}                  & \multirow{2}{*}{[-0.03, 0.01]} & \multirow{2}{*}{-0.01} \\

  \Wquart & 9.55 &  &  &  & 3.60 &  &  &  &  &  &  \\ 
  \arrayrulecolor{Gray}\hline

  \Wquart & 14.11 & \multirow{2}{*}{[-2.10, 0.43]} & \multirow{2}{*}{[-2.51, 0.83]} & \multirow{2}{*}{-0.85} 
  & 3.61 & \multirow{2}{*}{[-0.56, 0.12]} & \multirow{2}{*}{[-0.67, 0.24]} & \multirow{2}{*}{-0.22} & 
  \multirow{2}{*}{[-0.03, 0.00]} & \multirow{2}{*}{[-0.04, 0.01]} & \multirow{2}{*}{-0.01} \\

  \Wten & 13.59 &  &  &  & 3.47 &  &  &  &  &  &  \\ 
  \arrayrulecolor{black}\midrule

  \Iclick & 7.62 & \multirow{2}{*}{[-0.59, 0.64]} & \multirow{2}{*}{[-0.79, 0.82]} & \multirow{2}{*}{0.02} & 
  3.79    & \multirow{2}{*}{[-0.20, 0.11]}        & \multirow{2}{*}{[-0.24, 0.16]} & \multirow{2}{*}{-0.04} & 
  \multirow{2}{*}{[-0.02, 0.00]}                  & \multirow{2}{*}{[-0.02, 0.00]} & \multirow{2}{*}{-0.01} \\

  \Iget & 6.93 &  &  &  & 3.88 &  &  &  &  &  &  \\ 
  \arrayrulecolor{black}\midrule

  \Pleft & 8.33 & \multirow{2}{*}{[-0.65, 0.98]} & \multirow{2}{*}{[-0.91, 1.25]} & \multirow{2}{*}{0.16} & 
  3.72   & \multirow{2}{*}{[-0.03, 0.20]}         & \multirow{2}{*}{[-0.07, 0.24]} & \multirow{2}{*}{0.08} & 
  \multirow{2}{*}{{\bf [-0.03, -0.02]}}           & \multirow{2}{*}{{\bf [-0.03, -0.02]}} & \multirow{2}{*}{-0.02} \\

  \Pright & 7.47 &  &  &  & 3.33 &  &  &  &  &  &  \\
  \arrayrulecolor{black}\midrule

  \Dtop   & 13.73 & \multirow{2}{*}{{\bf [5.04, 6.67]}} & \multirow{2}{*}{{\bf [4.78, 6.92]}} & \multirow{2}{*}{{5.86}} & 
  4.59    & \multirow{2}{*}{{\bf [1.05, 1.24]}}         & \multirow{2}{*}{{\bf [1.02, 1.26]}} & \multirow{2}{*}{{1.14}} & 
  \multirow{2}{*}{{\bf [-0.02, -0.01]}}                 & \multirow{2}{*}{{\bf [-0.03, -0.01]}} & \multirow{2}{*}{{-0.02}} \\

  \Dbottom & 5.97 &  &  &  & 3.12 &  &  &  &  &  &  \\ 
  \bottomrule
\end{tabular}%
}
\caption{Impact of changes caused by varying video watch times (rows 1-3),
interaction mechanisms (row 4), recommendation selection strategy (row 5), 
and crawl depth (last row). Columns represent the mean node values observed
in each group for a particular characteristic, the 95\% and 99\% confidence
intervals for the measured effect sizes, and the mean effect size. Values in
bold indicate a statistically significant effect size at the corresponding
confidence level.}
\label{tab:rq3-results}
\end{table*}

\section{Computational Compromises} \label{sec:computation}

\para{Experiment setup.} In this section, we analyse the impact of three
compromises that may be made to save computational resources: 
(1) watching only a pre-determined fraction of each video in the recommendation
tree;
(2) using the {\tt driver.get(URL)} method of selenium rather than automating
user clicks on recommended videos; and
(3) performing low-depth and narrow-breadth audits.

\parait{Measuring impact of video watch times.}
To answer the question of whether audits need to `watch' videos to completion,
we gathered and analyzed four recommendation trees in which the audit `watched'
all videos to completion (\treeseta{w=100}), eight trees in which the audit
only `watched' videos to 50\% of their total duration (\treeseta{w=50},
\treesetb{w=50}), and four trees in which the audit only `watched' videos to
25\% of their total duration (\treeseta{w=25}). Both sets of audits used
the (\Tmain, \Smain) training set and seed.

\parait{Measuring impact of interaction mechanics.}
We gathered four recommendation trees where the audit actually located and
clicked the recommendations video links (\treeseta{click}) and four trees where
the audit simply identified the URL of the recommended videos and fetched the
video with a {\tt driver.get(URL)} command (\treeseta{get}). Both sets of
audits used the (\Tmain, \Smain) training set and seed.

\parait{Measuring the impact of crawl-breadth and -depth.}
We analyzed the characteristics of the leftmost and rightmost paths of all 96
recommendation trees gathered in this study (\treeseta{left} and
\treeseta{right}). These correspond to the paths obtained from only clicking
the top and bottom recommendation at each video, respectively.
We also analyzed the characteristics of the recommendations observed at depth
1 and 10 for all 96 trees obtained in this study (\treeseta{top} and
\treeseta{bottom}). 
 
Like before, in each of these analyses, we compute the means, 95\% and 99\%
confidence intervals associated with within-group differences, across-group
differences, and effect sizes.

\para{Results.} Our results are shown in \Cref{tab:rq3-results}. Notably,
besides configurations with varying crawl depth, none of our changes
yielded statistically significant differences in their measured recommendation
characteristics. This has several key implications for auditors.

\parait{Videos do not need to watched to completion.}
In all our audit configurations that varied video watch time fractions, there
was no statistical relationship between change in the characteristics of
recommended videos and the audit's configured watch fraction. 
This is a surprising finding that suggests even watching 10\% of a video
impacts the subsequent recommendations to no different extent as watching
100\%.
Upon further investigation, we discovered evidence showing that YouTube only
requires a watch time of 30 seconds for a `view' to be registered
\cite{growtraffic, tubics}. Based on these previous findings, we hypothesize
that this same 30-second watch threshold is also used to determine whether
a video should impact subsequent recommendations. 
Since the videos in our recommendation trees were much longer than 300-seconds
(with many being between 20-60 minutes long), even watching 10\% of the video
would register as a `view'.
This finding that videos do not need to be watched to any specific fraction
of completion, but rather to a fixed watch time threshold, presents
a promising (accuracy-independent) computational cost-saving avenue for future
auditors.

\parait{It is unnecessary to automate clicks on recommended videos.}
Our analysis showed no statistically significant differences between any 
recommendation tree characteristics observed in \treeseta{get} and
\treeseta{click}.
This suggests that using browser automation tools (\eg Selenium webdriver's
action chains \cite{actionchains}) to explicitly click on video links is
unnecessary. 
Without sacrificing on accuracy of audit inferences, this allows auditors to
replace a computationally expensive, high programmer overhead, and unreliable
approach to navigate to subsequent recommendations with the simple and reliable
approach of programming browsers to fetch specific URLs in the DOM.

\parait{Crawl depth impacts recommendation characteristics.}
Our analysis on the impact of crawl-depth yield statistically significant
results for all recommendation tree characteristics. Specifically, we notice
that nodes at the top of the recommendation tree generally appear to be more
significantly more popular, diverse, and less semantically similar to
recommendations at the bottom of the tree. 
This finding once again showcases the possibility of a strong recency bias that
impacts recommendations. 
Interestingly, we do not see statistically significant differences between the
highest- and lowest-recommended videos --- suggesting that auditors need to pay
specific attention to the depth of their crawls.

\para{Takeaways.}
Our analysis yields two significant computational cost-savings for researchers.
Specifically, finding that videos do not need to be watched to completion and
that clicking on videos causes no different outcomes than simply `getting' the
URL associated with the video reduces the computational and engineering
overhead associated with an audit.
In addition, our work highlights that different depths of a recommendation
tree could result in different recommendation characteristics. To account for
these effects, it is important that any inferences from an audit are
conditioned on the depth of the trees that were used.

\section{Related work} \label{sec:related-work}

%

\para{Audits of YouTube's recommendation system.}
This paper was inspired by a recent influx of YouTube audit research which
often showed contrary results. For instance, Lutz \etal \cite{lutz-2021}
provided evidence of the absence of a rabbit-holing effect while demonstrating
a mainstreaming effect for a variety of political ideologies. 
Other work \cite{ledwich-2019,munger-2022,hosseinmardi-2021,makhortykh-2020}
has also challenged the notion of rabbit-holing on YouTube and shown evidence
of recommendations swaying users towards mainstream and neutral content.
Contrary to these findings, Haroon \etal \cite{haroon-2022} provided evidence
that YouTube pushes users towards increasingly biased and radical political
content on `up-next' and homepage recommendations. 
These findings are complementary to another body of work
\cite{bryant-2020,ribeiro-2019,Tomlein-2021,kirdemir-2021,papadamou-2019,papadamou-2022}
which have argued that YouTube recommendations have promoted polarization in
the political, scientific, and health-related domains.
{Unlike these previous efforts, our goal is not to support or undermine
specific theories about YouTube's tendency to impact polarization. Rather, we
aim to uncover the possible reasons for these differences and provide
guidelines to avoid such confusion and contradictions within the auditing
community.}
More recently, Hussein \etal \cite{hussein2020measuring} showed how
demographics of a user profile altered recommendations from YouTube. In a study
focusing on YouTube's demonetization algorithm, Dunna \etal \cite{dunna-2022}
found evidence that the recommendation and demonetization algorithms were
linked. There are also numerous publications from Google describing the
recommendation algorithm used for YouTube. These have suggested the use of user
profiles, watch histories, video watch times, and click-through rates as
features in their content ranking algorithm \cite{zhao-2019, tang-2019,
fu-2016, covington-2016,  zhao-2015, brodersen-2012, davidson-2010}. 
These descriptions informed our choice of audit parameters. 

\para{Improving the reliability of crawler-based research.}
There have been similar efforts to ours in the Internet measurement community.
These have largely focused on facilitating more reliable and reproducible
research in the realm of Web measurement and privacy.
Yadav \etal \cite{yadav-2015} studied a set of open-source web crawlers and
showcased how each was suitable for different use cases. More recently, Ahmed
\etal \cite{suleman-2020} showed the impact that different crawlers had on
measurement and security research inferences. Along similar lines, Zeber \etal
\cite{zeber-2020} and Jueckstock \etal \cite{brave} also showed how the choice
of crawler and configuration could harm the repeatability of an experiment.
{Our work extends these efforts by identifying platform-specific audit
challenges.}

\section{Concluding Remarks} \label{sec:discussion}

\para{Limitations.} 
Fundamentally, our work is a best-effort study to
understand the impact of different audit methodological decisions on
recommendations gathered from YouTube --- one of the highest streamed video-hosting
platforms \cite{yt-rank}. While our audit has a YouTube-limited
scope, it helps pave way for other auditing studies across different video 
platforms.
Thus, our study is not without limitations. First, we ourselves are computationally
and economically limited and had to make decisions about crawl parameters to
explore. This impacted our ability to (1) perform exploration of more paths in
each recommendation tree; (2) conduct more than eight synchronized tree
explorations; and (3) explore recommendation trees to a greater depth. 
We take care to mitigate any incorrect inferences that might result from these
limitations by only performing like-for-like node- and position-dependent
comparisons and ensuring that any differences measured in our study account for
the general probabilistic nature of the recommendation algorithm by measuring
across-group differences and comparing them with within-group differences.
Second, there are latent effects that cannot be controlled from our external
vantage point which is effectively measuring a black-box system. We do our best
to identify several of these (\eg A/B testing, data center location,
measurement location, \etc) and attempt to counter each of them. However, it is
possible that unaccounted effects might still impact our results.
Finally, we acknowledge that our choice of a training set and seed video might
ultimately not be sufficient to observe all effects of interactions on the
recommendation system. Regardless, we provide useful data points for
consideration to a community grappling with a large number of contradictory
results.

\para{Conclusions.}
This work showcased the effect of audit configurations on the characteristics
of recommendation trees generated by them. 
Specifically, we showed that although training sets do have a statistical
impact on recommendations, their effects can be significantly dampened by
a `recency bias' in YouTube's recommendations (\Cref{sec:priors}). Therefore,
specific care needs to be taken when selecting videos to view in an audit. More
importantly, these decisions need to be disclosed and any audit inferences {\em
must} be conditioned on them.
Our analysis of different types of auditing profiles (\Cref{sec:dollars}) showed
that the expensive task of obtaining clean YouTube accounts would not yield
significantly different outcomes than simply maintaining the YouTube cookie for
the entire duration of an audit. Further, our findings also suggest that
account reuse can be possible by using the `clear history' feature provided by
YouTube.  
Finally, our analyses of various computational compromises in audits
(\Cref{sec:computation}) show that audits do not need to watch a specific
fraction of a video for it to impact subsequent recommendations (rather,
a preset threshold appears sufficient), challenging automation tasks such as
programming cursor clicks on videos do not need to be performed by auditors,
and that the depth of a crawl can impact characteristics of the recommendation
tree (and should therefore be used to condition any reported inferences from
audits).

\balance

\bibliographystyle{aaai21}
\bibliography{crawling}
\end{document}